\documentclass[nofootinbib,showpacs,notitlepage,prd,aps,longbibliography,twocolumn]{revtex4-1}
\newcommand{\be}{\begin{equation}}\newcommand{\ee}{\end{equation}}
\newcommand{\bea}{\begin{eqnarray}}\newcommand{\eea}{\end{eqnarray}}
\newcommand{\nn}{\nonumber}

\newcommand{\om}{\omega}
\newcommand{\la}{\lambda}\newcommand{\al}{\alpha}
\newcommand{\Ga}{\Gamma}

\newcommand{\pt}{\tilde{p}}
\renewcommand{\phi}{\varphi}
\renewcommand{\k}{\mathbf{k}}\newcommand{\x}{\mathbf{x}}
\newcommand{\Ref}[1]{(\ref{#1})}

\renewcommand{\kappa}{\varkappa}
\usepackage{epsfig}

\begin{document}
\title{Surface plasmon on graphene at finite $T$}
\author{M. Bordag\footnote{bordag@itp.uni-leipzig.de}}
\affiliation{ Leipzig University, Institute for Theoretical Physics,  04109 Leipzig, Germany}
\author{I.G. Pirozhenko\footnote{pirozhen@theor.jinr.ru}}
\affiliation{ Bogoliubov Laboratory of Theoretical Physics, Joint Institute for Nuclear Research and
Dubna International University, Dubna 141980, Russia}

\date{\small \today}
\begin{abstract}
Using the recently developed polarization tensor in (2+1) dimensions for the electronic excitations of graphene, we investigate the influence of temperature on the surface plasmons on graphene. We consider non-zero mass gap, but zero chemical potential. Plasmons may exist for both polarizations, TE and TM, of the electromagnetic field. For TE, the momentum region, where the dispersion function is real, appears bounded from below, whereas for TM it is bounded from above. We discuss the similarities of these features with those found previously in the case with non-zero chemical potential, but zero temperature.
\end{abstract}
%

\maketitle
\section{Introduction}
Electromagnetic waves travelling along a surface or an interface between different media, or along a thin sheet, are a well known phenomenon, especially on the surface of metals \cite{Raether1988b}. The amplitude of such waves decreases exponentially in the directions perpendicular to the interface. Polarization and dispersion of such waves depend on the properties of the medium behind the interface or of the sheet. Such waves are frequently called {\it surface plasmons}. However, there is no unique terminology. So, for instance, these waves are called sometimes simply 'plasmons'. This latter notion is mostly used for waves of density oscillations within a medium, a plasma half-room  for example,  together with their accompanying  electromagnetic fields. These are longitudinal, whereas the surface plasmons are transversal.

Surface plasmons may travel on graphene. These have been investigated since \cite{vafe06-97-266406,wuns06-8-318} in quite a number of papers
\cite{hwan07-75-205418,mikh07-99-016803,hans08-103-064302, herb08-100-046403,stau14-26-123201} and in \cite{bord14-89-035421,bord15-91-085038}. The electronic excitations in graphene are modeled by a Dirac equation. The basic method of investigation is the random phase approximation for the density-density correlation function. Another, equivalent but possible less intuitive,  method is the application of the methods known from relativistic quantum field theory. It rests on the description of the system, consisting of the electromagnetic field and the electronic excitations, as a quantum electrodynamics with photon in (3+1) dimensions and a fermion in (2+1) dimensions (with different speeds of light). Within this formalism, the problem to be solved consists of the calculation of the photon polarization tensor $\Pi_{\mu\nu}(k)$ as a one-loop Feynman diagram using standard methods. The fermion has the Fermi speed $v\sim c/300$ as its 'speed of light'. This way the theory has two Lorentz groups, one in (3+1) dimensions with the speed $c$ for the photon, and another one in (2+1) dimensions with speed $v$ for the spinor. These two interact through the usual minimal coupling in a consistent way.


The polarization tensor $\Pi_{\mu\nu}(k)$ in the form, allowing an analytic
continuation to the real frequency axis, was applied to the calculation of
optical properties of graphene in \cite{bord15-91-045037}, and to the calculation of the
Casimir effect in \cite{bord09-80-245406,fial11-84-035446,bord12-86-165429}. For the case of graphene in a magnetic field,
see \cite{gusy07-21-4611} and references therein. The relation of  these calculations to those using density-density correlation functions, was discussed in \cite{klim14-89-125407}. A characteristic feature to be mentioned is that the polarization tensor has, in the presence of a medium, i.e., with $T$ and/or $\mu$, two independent form factors. The results obtained using both methods and their relation to the van der Waals force were compared in \cite{klim13-87-075439}.

The polarization tensor with both form factors was (re-)derived in \cite{bord15-91-085038}, for $\mu\ne0$, $T=0$, where also its relation to the reflection coefficients on graphene was discussed in detail. These formulas were applied to the investigation of surface plasmons. The case with $\mu=0$, $T\ne0$ was considered in \cite{klim15-91-174501} with application to the large thermal contributions to the Casimir force on graphene.

In the next section we introduce the necessary notations. In section 3 we investigate the TE-plasmon, in section 4 the TM-plasmons and conclusions are given in the last section.

Starting from here, throughout the paper we use units with $\al=e^2/(4\pi)$ for the coupling and $v=v_F/c$ for the Fermi speed, and refer to $\al=1/137$ and $v=1/300$ as the physical parameters.

\section{Polarization tensor in plasmonic region}
Our notations follow \cite{bord15-91-085038}, where electrodynamics with polarization tensor was considered in detail. Reflection and transmission coefficients for the scattering of a incident plane electromagnetic wave off a graphene located at $z=0$ are
\be     r_{\rm TX}=\frac{-1}{1+Q_{\rm TX}^{-1}},\quad t_{\rm TX}=\frac{1}{1+Q_{\rm TX}},
\label{1}\ee
correspondingly  with TX$\to$TE for the transverse electric (TE) polarization and with TX$\to$TM for the transverse magnetic (TM) polarization of the wave. In terms of the form factors we get
\be\begin{array}{rclrl}
    Q_{\rm TE} &=& \frac{ 1}{2\eta}\left(\frac{\tilde{p}^2}{v^2k^2}\Pi^{00}(\tilde{p}) +\Pi_{\rm tr}(\tilde{p})\right),
\\[4pt]      Q_{\rm TM} &=& \frac{\eta}{2v^2k^2}\,\Pi^{00}(\tilde{p}),
\end{array}\label{2}\ee
where $\Pi_{\rm tr}(\tilde{p})=g_{\mu\nu}\Pi^{\mu\nu}(\tilde{p})$ is the trace of the polarization tensor.

The photon wave function  has plane waves proportional to $\sim\exp(-i\om t+i\k\x\pm i p z)$ (outside the plane, i.e., for $z\ne0$) with $\k=(k_1,k_2)$ (and $k=|\k|$) -- the momentum components parallel to the plane at $z=0$, and
\be     p=\sqrt{\om^2-\k^2},
\label{3}\ee
for the momentum perpendicular to the plane.
The argument of the form factors is the momentum
\be \tilde{p}=\sqrt{\om^2-(v k)^2},
\label{4}\ee
which involves the Fermi speed $v\sim 1/300$.

Surface plasmons appear as poles of the reflection coefficients, thus as solutions of the equations
\be    1+ Q_{\rm TX}=0.
\label{5}\ee
These solutions determine the frequency of the plasmons as function of the momentum,
\be     \om=\om_{\rm sf}(\k),
\label{6}\ee
i.e., the dispersion of the surface plasmons.
In order not to decay in time, these solutions must have real $\om$. Furthermore, these must have imaginary $p$ to decay in the direction perpendicular to the plane,
\be p=i\eta\equiv i\sqrt{-\om^2+k^2}.
\label{7}\ee
This implies a dispersion below the continuous spectrum,
\be v k<\om<k.
\label{8}\ee
The lower bound comes from the polarization tensor, which otherwise would not allow for solutions of \Ref{5} with real $\om$. In this way, it is just the second speed of light which opens a window for the existence of surface plasmons. We call this region the {\it plasmonic region}.

We represent the quantities $Q_{\rm TX}$, entering \Ref{1}, in the form
\bea               Q_{\rm TE} &=& \frac{ 1}{2\eta}\left(
            \frac{ \pt^2}{v^2k^2}(h_{00}+g_{00})
                +(h_{\rm tr}+g_{\rm tr})\right),
\nn\\[4pt]        Q_{\rm TM} &=& \frac{\eta}{2v^2k^2}(h_{\rm 00}+g_{\rm 00}),
\label{9}\eea
where
\be\label{10}\begin{array}{rclrcl}
   h_{00}&=&-\al\left(\frac{vk}{\pt}\right)^2\Phi(\tilde{p}), &
   \\[4pt]
    h_{\rm tr}&=&2\al \ \Phi(\tilde{p}) ,&
\end{array}\ee
are the zero temperature parts. These depend on a single function,
%
\bea &&\Phi(\tilde{p}) \label{11}\\
&&=\left\{ \begin{array}{ll}
\frac{2}{\pt}\left(2m\pt-(\pt^2+4m^2){\rm arctanh}\frac{\pt}{2m}\right),&(\om<\om_s )     \\
    \frac{2}{\pt}\left(2m\pt-(\pt^2+4m^2)\left({\rm arctanh}\frac{2m}{\pt}+\frac{i\pi}{2}\right)\right),&(\om>\om_s)   \end{array} \right.    \nn
\eea
only. This function has a threshold at
\be \om_s=\sqrt{(v k)^2+(2m)^2}.
\label{12}\ee
Above, i.e., for $\om>\om_s$, electron-hole creation is possible and the polarization tensor has an imaginary part. The resulting plasmons would not be stable.
This way,
\be \om<\om_s,
\label{13}\ee
poses another condition on the plasmonic region.

The temperature dependent part in \Ref{9} was derived in \cite{bord15-91-045037}. Changing notations to those used in this paper, it is given by an integral,
\be g_{X}=\frac{8e^2}{2\pi}\int_0^\infty dq\frac{q}{\Ga}\frac{1}{e^{\Ga/T}+1}
\left(1+\frac12\sum_{\la=\pm1}\frac{M_{X}}{N}\right),
\label{14}\ee
where '$X$' stands either for '$00$' or for 'tr'. Further notations are
\bea    \Ga&=&\sqrt{q^2+m^2}, ~~a=2vkq, \nn\\
        Q&=&-\pt^2-2\la \om\Ga  ,  ~~N= {\rm sign}(Q)\sqrt{Q^2-a^2} ,
\nn\\
M_{00}&=& 4\Ga^2 +\pt^2 +4\la \om\Ga ,~~
M_{\rm tr}= \pt^2+4m^2.
\label{15}\eea
These functions are real in the plasmonic region.

\begin{figure}[t]
\centerline{\epsfig{file=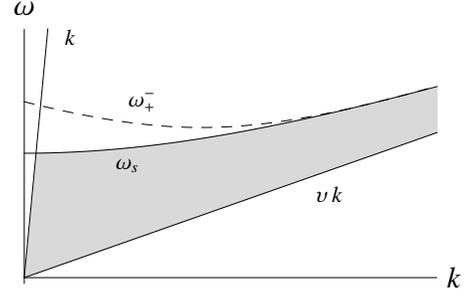,width=6cm}}
\caption{The plasmonic region defined by $v k<\omega<{\rm Min}(k,\omega_s)$. The curve $\om_+^-$ is shown for $q=1$, $m=1$, $v=1/30$.}
\label{regions}
\end{figure}

It should be mentioned, that the denominator, $N$, is real and non-zero in the plasmonic region. In fact, it is real in a larger region defined by $\om<\om_+^-$, where $\omega_+^{-}=\sqrt{k^2 v^2-2 k q v+m^2+q^2}+\sqrt{m^2+q^2}$ ($q$ is the integration variable in \Ref{14}) is a solution of $N=0$, and $\om<\om_1^-$ holds.
The plasmonic region is shown in Fig. \ref{regions}. It does not depend on $\alpha$.

In this way, all quantities entering Eqs. \Ref{5}, are defined, and in the next sections we investigate their solutions numerically.
\section{The TE plasmon}
\begin{figure}[t]
\centerline{\epsfig{file=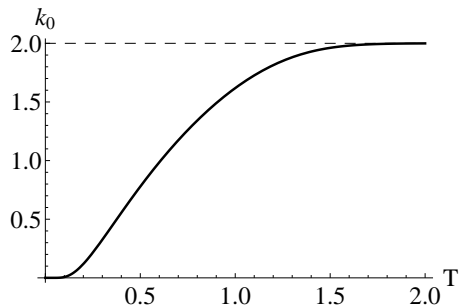,width=6cm}}
\caption{The starting point $k_0$ of the TE surface plasmon as a function of temperature $T$ for $\alpha=1/137$, $v=1/300$. The dashed line is  at  $k=2m/\sqrt{1-v^2}$.}
\label{TEk0}
\end{figure}

We start with the investigation of the TE plasmon.
At zero temperature the TE plasmon  solution was discussed in \cite{bord14-89-035421}.  It exists for all k. At finite $T$, this changes and the dispersion has a starting point, $k_0$. It is shown in Fig.~\ref{TEk0}  as a function of temperature.  Here $T$ and $k$ are given in the units of mass. For example, at $m=0.025$ eV we have $T=1$, which corresponds to room temperature.

\begin{figure}[h]
\epsfig{file=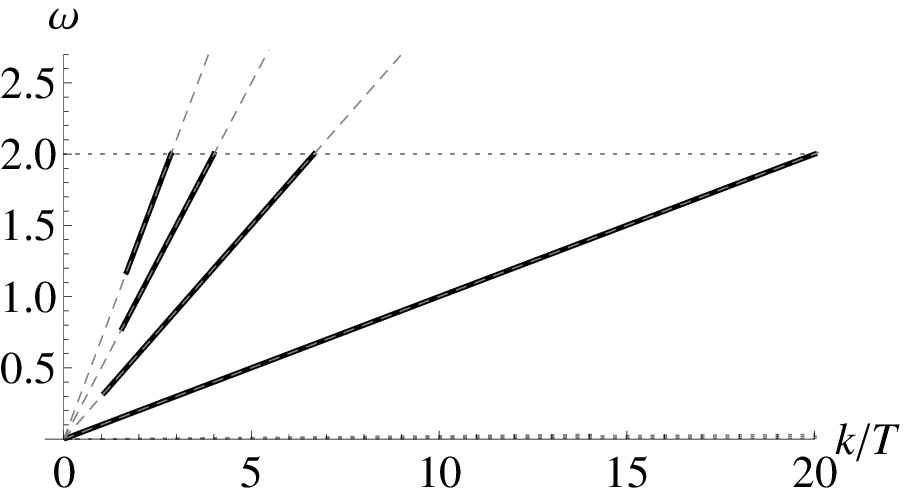,width=6cm}
\epsfig{file=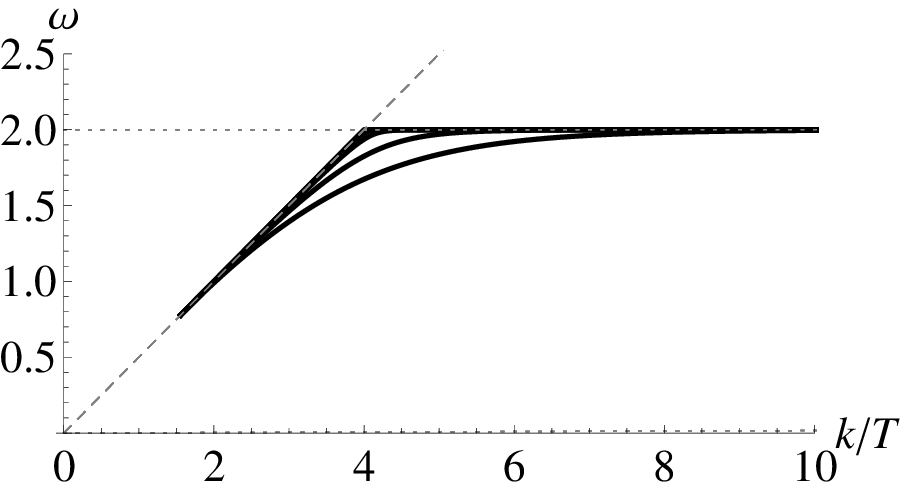,width=6cm}
\caption{TE surface plasmon frequency as a function of $k/T$.
Top: The dispersion of TE plasmon for physical values of the parameters  and for several values of the temperature,
$T=0.1, 0.3, 0.5, 0.7$ (from bottom to top). Bottom: TE plasmon frequency at $T=0.5$ for different values of the coupling constant, $\alpha=1/137, 1/30, 1/10, 1/5, 1/2, 1$ (from top to bottom).}
\label{TE_T}
\end{figure}

The starting point of the dispersion curve is on the border of the continuous spectrum, $\omega=k$, and the endpoint approaches  $\omega=\omega_s$.
Fig.~\ref{TE_T}(left) shows the dispersion relations for TE surface plasmons   for different temperatures and with $\alpha=1/137$, $v=1/300$.  The solutions are represented by nearly straight lines, whose lengths decrease with increasing  temperature. Represented in the $(k,\omega)$-plane, they all stick together. Therefore we represented them in the $(k,\omega/T)$-plane.

As mentioned, these lines become shorter with increasing temperature. There is a largest temperature when the length shrinks to zero. This happen when the beginning point reaches the intersection of the border of the continuous  spectrum, $\om=k$, with the threshold, $\om=\om_s$.
The  numerical solution of Eq. \Ref{1}, with $\omega=k$ and simultaneously $\omega=\omega_s$, gives $T_{max}\sim 3.76$.
However, already for $T>2.25$ the TE plasmon is invisible on the plot.
As known from $T=0$, the curves have a knee for $k>2m/\sqrt{1-v^2}$ (see \cite{bord14-89-035421}). For coupling stronger than its physical value, the knee at the end of the curves becomes visible, see Fig.~\ref{TE_T} (right).

\section{The TM plasmon}
\begin{figure}[t]
\epsfig{file=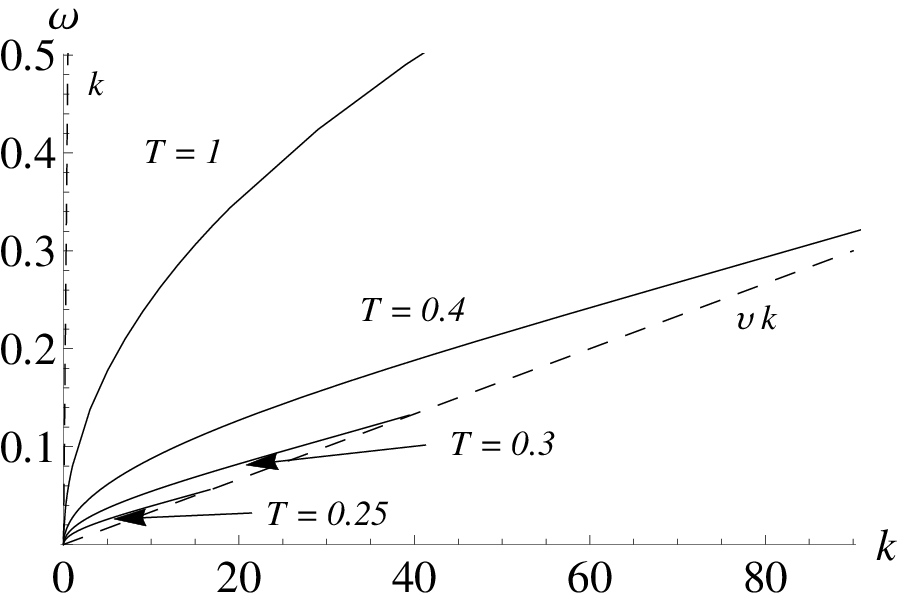,width=6cm}
\epsfig{file=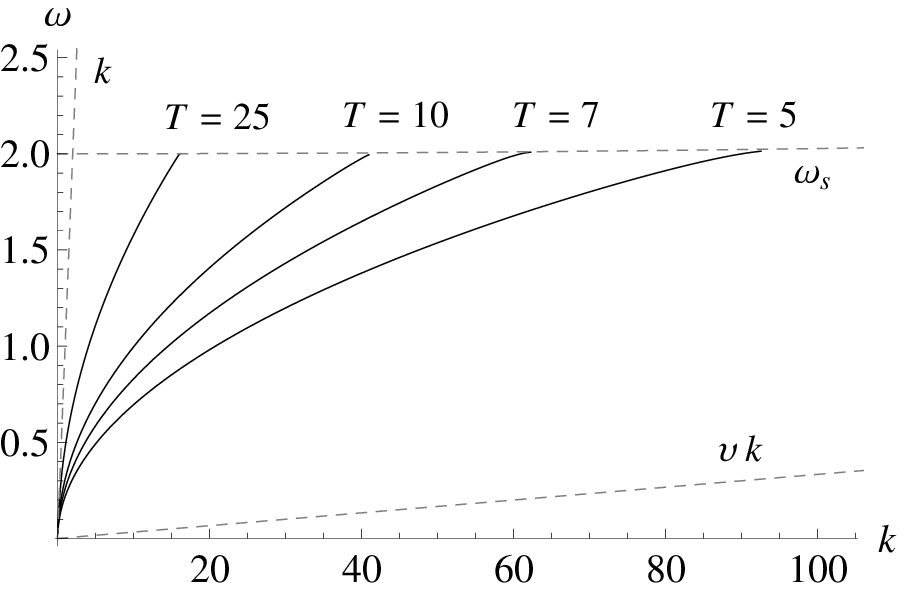,width=6cm}
\caption{The dispersion of the TM plasmon for physical values of the parameters and $m=1$ for several values of the temperature, low $T$ in the upper panel, higher $T$ in the lower panel. The dashed lines are $\om=v k$ and $\om=k$, the dotted line is the threshold $\om=\om_s$.}
\label{TM_T}
\end{figure}

The TM plasmon is the solution of the equation $1+Q_{TM}=0$, \Ref{5}.  It is absent for $T=0$.
The dispersion curve starts  at $k=0$. The curve has an endpoint whose position   depends on the temperature $T$.
It can be calculated analytically,
\begin{equation}
k_{fin}=\frac{3 m}{2 \alpha \sqrt{1-v^2}}\left[-1+\sqrt{1+\frac{16}{3 m}\frac{\alpha^2(1-v^2)}{v^2} \mathcal{F}(T)}\right]
\label{16}\end{equation}
with
\be
\mathcal{F}(T)=\frac{\pi^2 T^2}{3 m}+\frac{2 T^2}{m}\mbox{PolyLog}[2,-e^{m/T}]+T\ln[1+e^{m/T}].
\label{17}\ee
At zero temperature $\mathcal{F}(T)$ vanishes, the endpoint is on $k=0$ and the TM plasmon disappears.
The dispersion of the TM plasmon is shown on Fig. \ref{TM_T}.

With increasing  $T$ the curves may partially go beyond the region allowed for real solutions.
Thus, two real branches of one solution appear, separated by a gap where the latter is not real.
At relatively high temperature ($T>2.75$), the real solution disappears at the boundary $\omega=\omega_s$. It re-appears as second branch  at large momenta, $k<k_{end}$. For $T=2.75$,  $k_{end}/m\approx3176$, which is far above the applicability region of the Dirac model.
In this sense, for high temperature   only the low momentum branch terminating  at the threshold $\omega=\omega_s$, Fig. \ref{TM_T} (right), is physical.

\section{Conclusion}
\begin{figure}[t]
\epsfig{file=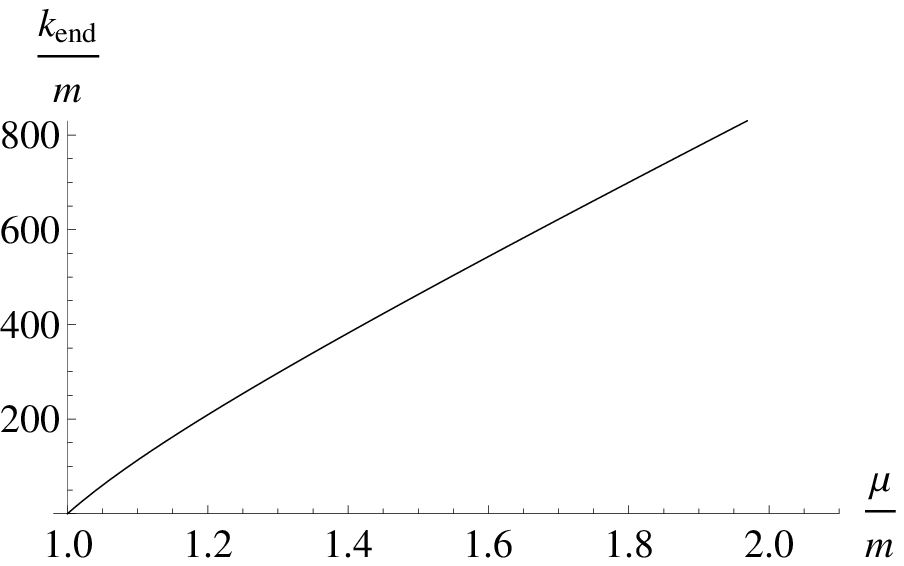,width=6cm}
\epsfig{file=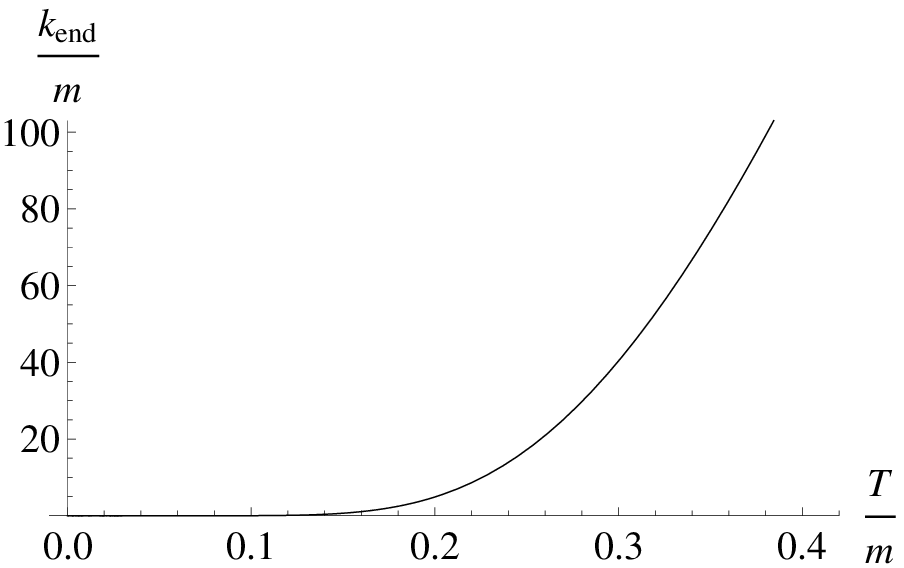,width=6cm}
\caption{The endpoints of the dispersion corves of the TM plasmons for physical values of the parameters. In the upper panel for $T=0$ $\mu\ne0$ and in the lower panel for $T\ne0$ and $\mu=0$
.}
\label{endpoints}
\end{figure}

%
In the foregoing sections we investigated the surface plasmons on graphene at finite temperature and non-zero mass (gap parameter). A common feature is that these plasmons may exist for both polarizations, TE and TM. Examples are shown in Figs.\ref{TE_T} and \ref{TM_T}. Thereby, with increasing temperature, the TE dispersion curve has a starting point $k_0$, see Fig. \ref{TEk0}, whereas the TM dispersion has an endpoint, see Fig. \ref{endpoints}. As a result, for TE there exists a temperature, above which there is no solution, whereas in the TM case the solution disappears for low temperature.

It is interesting to observe similarities with the case of chemical potential $\mu$, which was considered in \cite{bord15-91-085038} at $T=0$. First of all, the plots for the dispersions look similar, see for TE Fig. 5 in \cite{bord15-91-085038} and Fig. \ref{TE_T} here, and for TM Fig. 3 in \cite{bord15-91-085038}  and Fig. \ref{TM_T} here. Also, for TE we have in both cases a starting point, see Fig. 4 in \cite{bord15-91-085038}  and Fig. \ref{TEk0} here. In TM polarization there are endpoints in both cases, which have even analytic expressions, eq. (59) in \cite{bord15-91-085038}  and \Ref{16} above. The corresponding plots are in Fig. \ref{endpoints}. These show a similar behavior, i.e., starting in $k_{\rm end}=0$ and growing up with the corresponding parameter, $\mu$ resp. $T$.It would be interesting to investigate this similarity in more detail by considering surface plasmons with non-zero both, $T$ and $\mu$.

Another interesting point would be the extension of the above methods to complex solutions, i.e., to resonances which can be viewed as plasmons with finite life time. These would be extensions of the dispersion curves beyond their endpoints. We mention that the extension beyond the starting points would give a solution in the continuous spectrum. These would have a wave function not decreasing in the direction perpendicular to the sheet and would be at best weakly localized at the sheet.

Another application could be the question on whether the plasmons on graphene will play the same role as with plasma sheets with respect to the van der Waals forces between two sheets which was investigated in \cite{intr05-94-110404} and \cite{bord06-39-6173}.


\end{document}